# Building Trustworthy Multimodal AI: A Review of Fairness, Transparency, and Ethics in Vision-Language Tasks

Mohammad Saleh*, Azadeh Tabatabaei

Department of Computer Engineering, University of Science and Culture, Tehran, Iran;

smohamad82@gmail.com,  a.tabatabaei@usc.ac.ir

## ABSTRACT

Objective: This review explores the trustworthiness of multimodal artificial intelligence (AI) systems, specifically focusing on vision-language tasks. It addresses critical challenges related to fairness, transparency, and ethical implications in these systems, providing a comparative analysis of key tasks such as Visual Question Answering (VQA), image captioning, and visual dialogue. Background: Multimodal models, particularly vision-language models, enhance artificial intelligence (AI) capabilities by integrating visual and textual data, mimicking human learning processes. Despite significant advancements, the trustworthiness of these models remains a crucial concern, particularly as AI systems increasingly confront issues regarding fairness, transparency, and ethics. Methods: This review examines research conducted from 2017 to 2024, focusing on forenamed core vision-language tasks. It employs a comparative approach to analyze these tasks through the lens of trustworthiness, underlining fairness, explainability, and ethics. This study synthesizes findings from recent literature to identify trends, challenges, and state-of-the-art solutions. Results: Several key findings were highlighted. Transparency: The explainability of vision language tasks is important for user trust. Techniques, such as attention maps and gradient-based methods, have successfully addressed this issue. Fairness: Bias mitigation in VQA and visual dialogue systems is essential for ensuring unbiased outcomes across diverse demographic groups. Ethical Implications: Addressing biases in multilingual models and ensuring ethical data handling is critical for the responsible deployment of vision-language systems. Conclusion: This study underscores the importance of integrating fairness, transparency, and ethical considerations in developing vision-language models within a unified framework.

*Keywords*— VQA, Ethical Implications, Trustworthiness, Debiasing; Explainability, Image Captioning, Visual Dialogue.

## I. Introduction

Computer Vision and Natural Language Processing have advanced significantly [1], surpassing human performance on various tasks [2, 3, 4]. The strengths of these algorithms and capabilities of autonomous systems underscore the importance of integrating diverse fields of knowledge to develop intelligent cross-modal solutions [5]. A Vision-language task involves combining visual and textual data to perform tasks that require understanding the relationship between these two modalities. One example is visual captioning, which generates meaningful language descriptions based on visual information [1].

This area of study is evolving rapidly, making it essential to explore recent trends, breakthroughs, and latest methods across various domains [5, 6, 7]. Numerous review papers have outlined these tasks, their key components, and the impact of recent technologies on them [6, 7]. This paper presents an analytical study of Vision-language tasks, focusing on the key challenges of trustworthy AI systems: transparency, fairness, and ethics. These challenges are crucial for modern AI systems, as a growing body of literature increasingly acknowledges their significance in artificial intelligence research.

Table 1 provides an overview of the most relevant studies related to our research. Many studies not focused on Vision-language tasks continue to address the aforementioned issues of trustworthiness in AI, such as fairness [11, 12, 13, 14, 15, 16], transparency [17, 18], and ethics [9]. Others have attempted to tackle the associated challenges of vision-language tasks [19], but have not specifically addressed the principles of trustworthiness. Furthermore, some reviews have focused on distinct Vision-language tasks [8].

This is the first review to examine Vision-language tasks through the lens of trustworthiness. Moreover, because the chosen Vision-language tasks are interconnected, analyzing them collectively will yield new insights. We begin by introducing our proposed taxonomy, outlining core Vision-language tasks, including Visual Question Answering, visual captioning, and dialogue. Next, we detail our comparative approach to these tasks and review recent advancements in Vision-language research, highlighting state-of-the-art findings for each challenge. Ultimately, this study provides insights to help developers create more trustworthy vision-language systems.

Table 1. The most related review papers

| Ref. | Year | coverage range | Task | Fairness | Transp. | Ethics. | Main theme |
|---|---|---|---|---|---|---|---|
| **Ours** | **2025** | **2017-2024** | **Vision Language Tasks** | ✓ | ✓ | ✓ | **Trustworthiness in VL Tasks** |
| [11] | 2025 | 2019-2024 | - | ✓ | - | - | LLMs & VLMs in general |
| [12] | 2024 | not mentioned | - | ✓ | - | - | Fairness in LLMs |
| [17] | 2021 | not mentioned | - | - | ✓ | - | ML interpretability methods |
| [19] | 2024 | not mentioned | Vision Language Tasks | | | ✓ | Impact of LLMs on VL tasks |
| [9] | 2024 | 2021-2024 | - | - | - | ✓ | Review about LLMs |
| [8] | 2022 | not mentioned | Visual Dialogue | - | - | - | Visual dialogue systems |

## I. Vision-language Tasks

Individuals encounter a vast array of information through different sensory modalities, and the human brain has evolved to effectively interpret these stimuli to understand the Environment [20]. Vision is very important for how we perceive things, while language is essential for communication. A multimodal AI system needs to accurately and efficiently manage these different types of information [21]. For instance, computers might achieve this by finding the most relevant images based on a text query, or by explaining the content of an image in natural language.

It is worth noting that vocal cues are important in how we perceive trustworthiness but are not as important as facial cues. This means that how an AI looks is more important [22]. Facial cues have a stronger influence on our perceptions of trustworthiness. As a result, users are more likely to rely on visual information when evaluating the trustworthiness of an AI [23]. The following sections give a concise overview of fundamental Vision-language tasks such as VQA, Visual captioning, and Visual Dialogue. Fig. 1 shows various Vision-language tasks that frequently have a lot in common.

### A. Visual Question Answering

Humans easily identify the surrounding objects and locate their position in the environment. We also infer the relationship between objects and recognize existing activities. Additionally, we can answer any desired questions about an image [24].

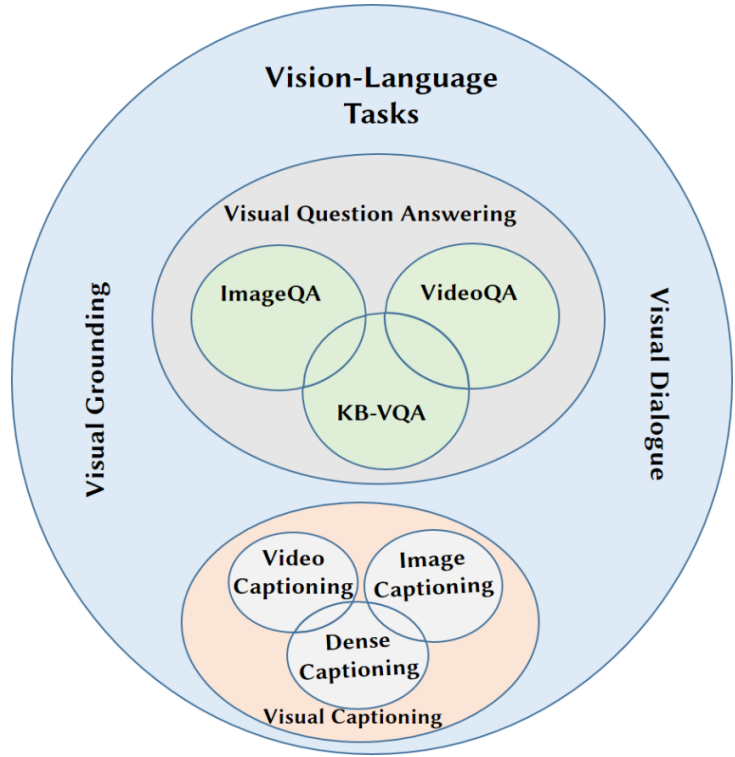

Figure. 1. Here, we highlight some core tasks in vision language along with their important subtasks. At the top, you will notice three key areas within VQA: image question answering, video question answering, and KB-QA. Down below, we highlight three types of Visual Captioning: Image Captioning, Video Captioning, and Dense Captioning. This Figure shows how two other main areas in Vision-language Research—Visual Reasoning and Visual Grounding—are connected to the tasks we have chosen.

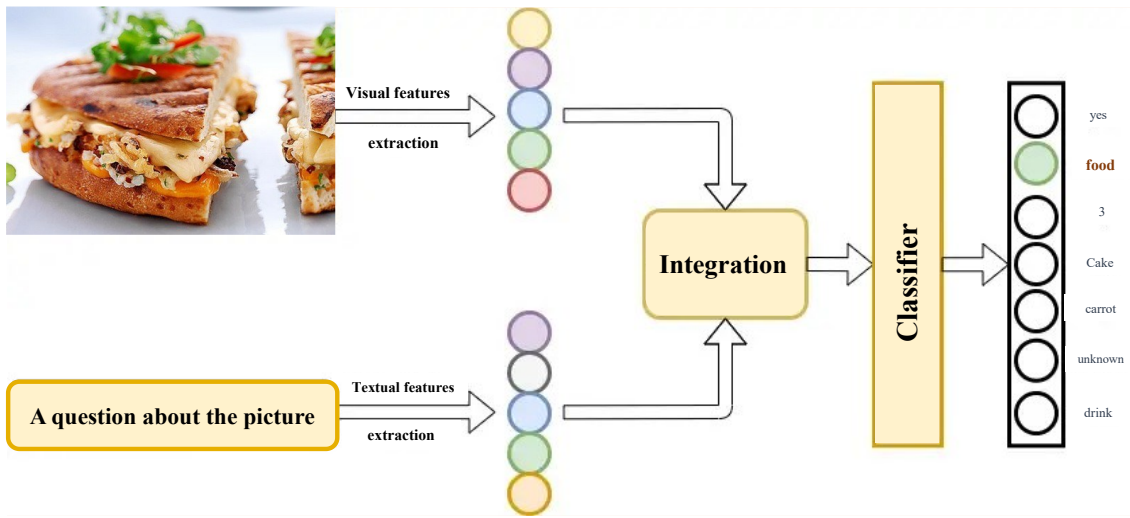

Figure. 2. Basic architecture of Image Question Answering models.

The ability to answer questions about content has long been recognized as the most prominent aspect of human perception.

The ability of a machine to answer questions about what it sees is known in the form of a Visual Question Answering problem [25], a multi-disciplinary research problem that combines natural language processing, computer vision, and knowledge-based reasoning [26]. In its simplest form, a system is given an image, a natural language question related to the image, and the model must respond in natural language that appropriately addresses the inputs [25]. (See Fig. 2.)

In recent years, many researchers have sought to tackle the problems and challenges in this field and have achieved numerous successes [27]. The VQA task has long served as a benchmark for evaluating the reasoning capabilities of AI systems and can be divided into several subcategories. Image Question Answering (ImageQA) involves understanding a single image and answering questions about it, while Video Question Answering (VideoQA) requires comprehending a sequence of images (a video) and responding to questions that involve temporal context [28, 29]. Knowledge-based VQA [30] (KB-QA) entails answering questions about an image or video by leveraging external knowledge from a knowledge base, such as Wikipedia or a structured database [31].

*B.* **Visual Dialogue**

Visual Dialogue involves an Artificial Intelligent agent engaging in a meaningful conversation with humans about visual content in natural language [8]. The task is to answer follow-up questions about an image, based on the given image and a dialogue history. Combining language and vision, it enhances AI's understanding of context, allowing for more intuitive interactions with users, and benefiting applications like customer service.

Visual Dialogue focuses on sequential questions and answers in a conversational format, which is important for developing AI systems that can effectively communicate and interact in real-world scenarios [32], such as with robots or virtual assistants. As technology advances, the ability of AI to understand and converse about visual content will greatly improve human-robot interaction [8]. This area of study has significantly contributed to the development of modern conversational chatbots [33].

*C.* **Visual Captioning**

Visual Captioning involves describing visual content in natural language, using a visual understanding system and a language model to create meaningful and grammatically correct sentences [34]. This task is illustrated in Figure 3. Visual Captioning can be used for a variety of purposes, such as adding metadata to images or making images more accessible to visually impaired people, automatic image indexing, and improving Content-Based Image Retrieval across various domains [35, 36].

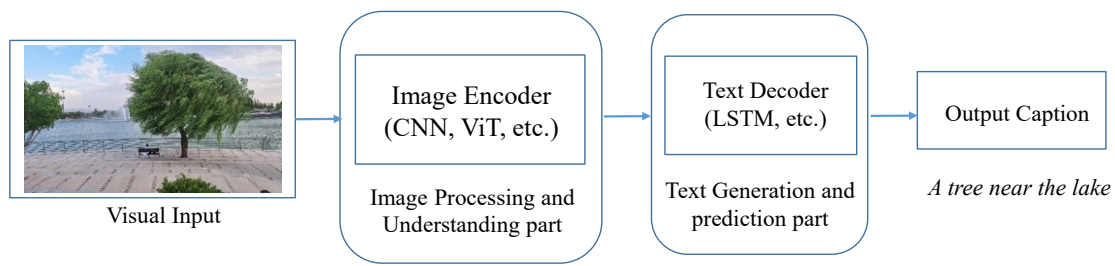

Figure. 3. Basic architecture of Image Captioning Models

A single sentence often fails to capture the rich content of images and videos, leading to the proposal of the Dense Captioning task, which generates multiple sentences for various detected object locations [37].

Visual captioning, closely related to VQA, has significantly contributed to the development of various VQA systems [38, 39] and has inspired multiple joint embedding VQA architectures [40].

*D.* **Other Tasks**

One of the most important Vision Language tasks is "Visual Reasoning", which focuses on understanding relationships and interactions within images [41]. This goes beyond merely answering questions about images (as in VQA) or engaging in dialogues about them (like Visual Dialogue).

While VQA specifically examines the interaction between visual data and natural language questions, visual reasoning is the foundational capability that enables these interactions [42]. The reasoning in VQA is a specific instance of the more general Visual Reasoning tasks, which include diverse decision-making and problem-solving relying on visual information. Therefore, VQA is a specialized application within the broader domain of Visual Reasoning [41, 42, 43].

II. **Methodology**

Trustworthiness is one of the important research directions of today's AI systems [44]. It suggests systems should work as expected while being safe and ethically responsible [44]. Trustworthy AI requires the fulfillment of several principles, including transparency, fairness, and ethical implications.

Transparency and Fairness are important elements shaping the ethical deployment of AI systems in practical settings [45]. Transparency significantly enhances user trust by enabling stakeholders to understand AI decision-making processes [13]. Studies show that when AI systems display Transparency [46], users are more likely to accept their recommendations, thereby promoting ethical practices.

For our study, we selected three interrelated fundamental vision-language tasks. VQA, Image Captioning, and Visual Dialogue. These tasks are depicted in Figure 1. The research papers we chose, published between 2017 and 2024, focus specifically on these three vision-language tasks. We employ a comparative approach to analyze these tasks through the lens of trustworthiness, emphasizing issues like fairness, explainability, and ethics.

The main objective of this study is to qualitatively assess the progress in trustworthy research related to these tasks and to identify future directions. We selected only the most influential and highly cited journal papers from ACM, Springer, and IEEE publishers, as well as from top-tier conferences such as CVPR, AAAI, and ICCV. Each task and issue has specific keywords that yield the most relevant results. For instance, terms like "fairness," "bias," and "debiasing methods" address fairness issues, while "explainability" pertains to transparency, and "ethics" and "ethical implications" cover ethical concerns. These keywords are most effective and should be combined with the tasks' specific names: VQA (Visual/Image Question Answering), Image Captioning, and Visual Dialogue.

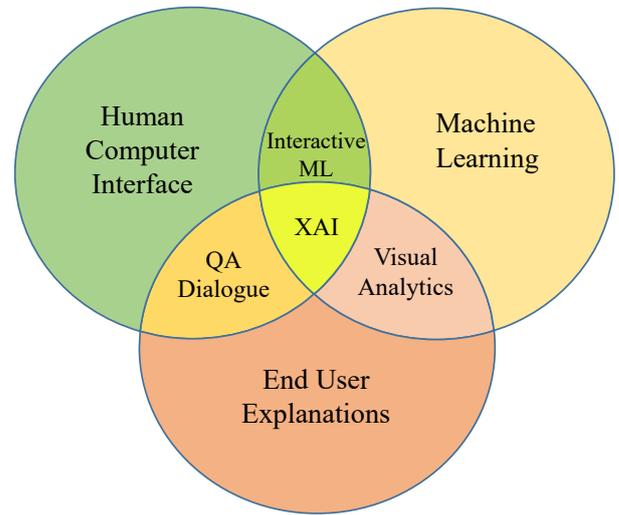

Figure. 4. Explainability in AI

The trustworthiness issues selected for study are interconnected, and our investigation into vision-language tasks can be divided into three main sections: A. Transparency, B. Fairness, and C. Ethical Implications.

*A.* **Transparency**

System transparency is closely tied to its explainability. Explainable AI (XAI) is a fascinating aspect of today's AI systems [47]. This field of study has become important in many areas, such as Finance and Medical Diagnosis [42, 47, 47, 48, 49, 50, 51, 52, 53]. Fig. 4 shows the relationship between Machine Learning (ML), Human-Computer Interaction (HCI), and end-user explanations. Fig. 5 presents our proposed research taxonomy. In Vision-language tasks, explainability refers to understanding and interpreting how a model makes decisions when processing visual data [54]. Explainability is essential for trustworthiness in Vision-language tasks since the clarity of a system's information strongly influences user trust [55]. Table 2 shows explainability methods for Vision-language tasks vary widely in their advantages and limitations.

*1)* **Explainability in VQA**

A VQA model should be able to provide facts or explain how it arrived at a given conclusion. If, during inference, the user can understand the logical flow from the input data being processed to the answer output, the model is considered explainable [17]. Large unified architectures [56], as well as multi-modal LLMs [57], have significantly improved the accuracy and generalization capacities of VQA models at the expense of model explainability. Several studies have suggested explainable architectures and addressed model explainability in VQA. However, modern VQA models are often viewed as a network of black-box modules or a black box itself, changing the field of VQA explainability.

*2)* **Explainability in Image Captioning**

Recent research on explainability in image captioning has aimed to clarify model decisions. Elguendouze et al. [58] used latent space perturbations to identify key components and compare explanation methods. Beddiar et al. [47] developed an explainable module for medical captioning, leveraging self-attention to link visual and semantic features. The attention mechanism is often used to create heatmaps that highlight areas in images that correspond to predicted captions [59].

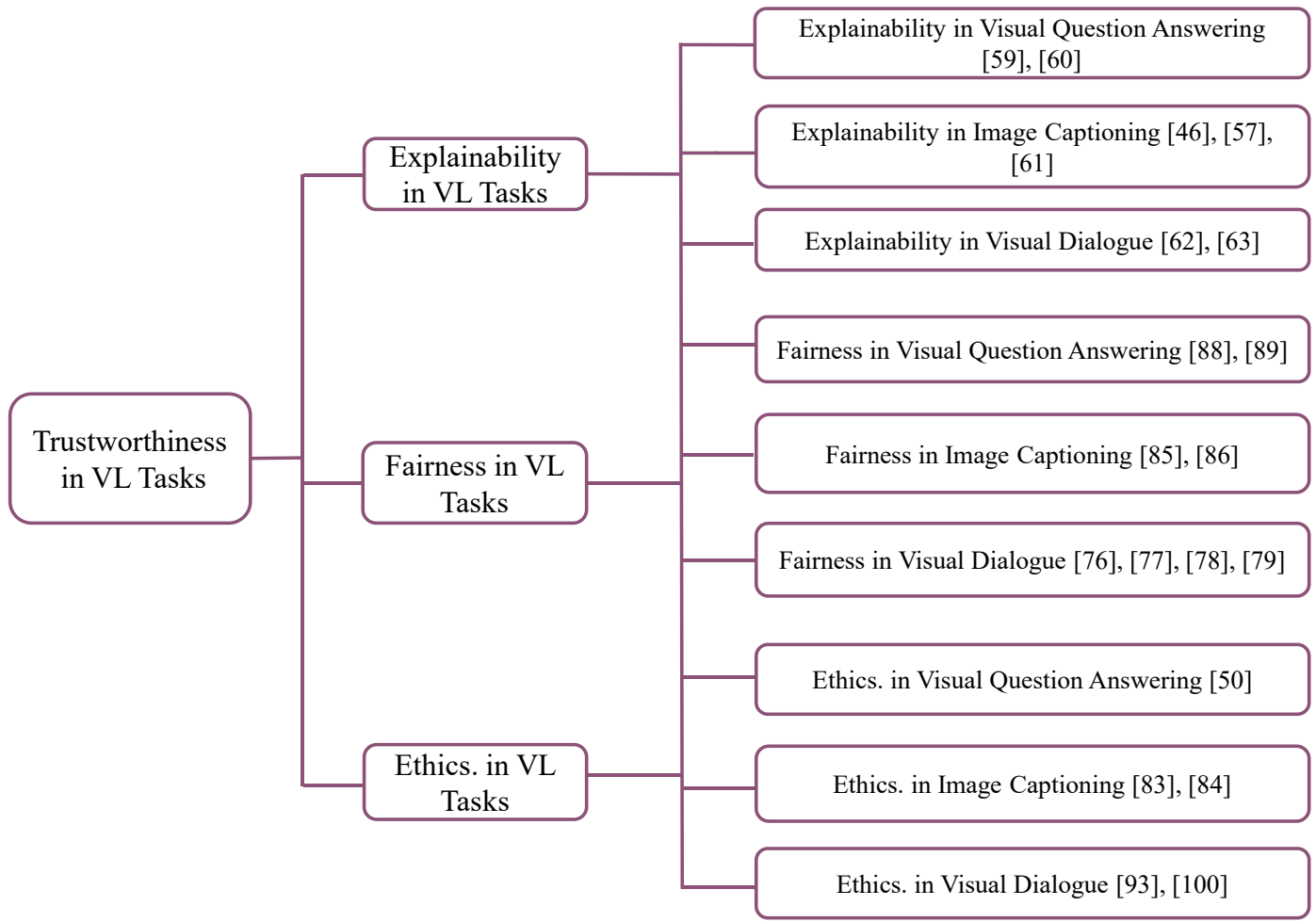

Figure. 5. Proposed taxonomy of Trustworthiness in the Vision-language tasks

Table 2. Research on Explainability of Vision-language tasks.

| Ref. | Task | Method | Advantage | Limitations |
|---|---|---|---|---|
| [60] | Visual Question Answering | Explanations directly linked to user expected answer | Telling why the answer to a question is P in contrast to F | Restricted complexity of the scenes |
| [61] | Visual Question Answering | Variational Causal Inference Network | Understanding how visual elements influence responses. | Struggle with narrative complexity in videos |
| [58] | Image Captioning | Analyzing latent space of NA Architecture | Shows which image parts influenced specific caption words | Visual part of VC is more decisive |
| [62] | Image Captioning | Link between image regions & captions | Evaluation on MSCOCO & Flicker30k | The reason behind generated captions? |
| [47] | Image Captioning | Self-attention computes word importance | Interpretation of encoder–decoder | Low quality descriptions |
| [63] | Visual Dialogue | Deconfounded Learning | Interactive mechanism | Limited labeled data & spurious correlations |
| [64] | Visual Dialogue | A novel data structure called Conversation memory | It holds information that is incrementally conveyed in the conversation | Reliance on visual input |

Han et al. [62] created a model that visually connects image regions to words. Al-Shouha & Szücs [65] proposed a segmentation-based explanation method to enhance trust.

*3) Explainability in Visual Dialogue*

Visual Dialogue models must provide explanations that adapt to ongoing conversations, adding complexity to the task. Explainability in these systems guarantees clear and comprehensible explanations for the decisions or responses made, often utilizing visual or textual formats. Deconfounded learning, as noted by [63, 66], significantly enhances Vision-language explanations and incorporates interactive mechanisms that elevate user feedback and boost the performance of dialogue systems.

Shen et al. [67] propose a data filtering method for open-domain dialogues that recognizes not-to-be-trusted training samples by linearly combining seven dialogue attributes for a quality measure. These initiatives enhance the transparency of dialogue systems.

Moreover, dialogue systems can enhance the explainability of AI applications independently [64, 68]. Danry et al. [69] present AI-framed Questioning, a concept that allows users to evaluate the logical validity of information. This approach enhances explainability, illustrating a future where AI agents collaborate with and challenge humans, rather than simply dictating beliefs or actions.

*B. Fairness*

Fairness refers to principles and techniques designed to prevent models from reinforcing biases or discriminating against specific individuals or groups [70]. The objective is to ensure that the models generate fair results for everyone, regardless of their background or characteristics [71]. Bias mitigation algorithms aim to improve fairness by modifying the training data [72], changing the learning process [73], or adjusting the final predictions [74]. This section will study this issue in selected Vision Language tasks. Table 3 highlights the most recent studies on fairness in Vision-language tasks.

*1) Fairness in VQA Models*

VQA models can be biased, producing incorrect or unfair results. Park et al. [75] Introduced a model that predicts equitable answers to sensitive questions while maintaining overall performance. In VQA, bias can be unintentional due to relying too heavily on one modality of the training data [76]. Based on the two types of modalities in these kinds of tasks, the model could have language and/or visual biases.

*2) Fairness in Visual Dialogue*

Research on fairness in Visual Dialogue focuses on ensuring that AI systems treat all users equitably, identifying and mitigating biases in visual dialogue systems, particularly those related to race, gender, and other demographic factors. This includes using diverse datasets for training and implementing fairness-aware algorithms [77, 78].

Research suggests that although bias mitigation techniques have the potential to decrease unfairness by as much as 23%, they may concurrently result in an approximate 9% decline in accuracy [79]. However, methods such as FairCLIP [80] demonstrate that this trade-off can be effectively managed.

*C. Ethical Implications*

*1) Ethical Implications in VQA*

Biased language models in VQA systems can create ethical concerns [81]. They can cause differences in performance and reinforce harmful stereotypes [82]. Furthermore, biases in multilingual language models can lead to inconsistent performance across languages, revealing hidden preferences and ignoring the needs of less-supported languages [83]. As VQA technology develops, it is important to address these ethical considerations to foster trust and ensure equitable access for everyone [25, 51].

*2)* **Ethical Implications in Image Captioning**

The advancement of Image Captioning algorithms highlights important concerns about biases and ethical considerations that must be addressed. Recent studies have shown that captioning systems can exhibit biases related to data, models, or both [84]. These biases may include gender, racial, and intersectional biases, which affect the captions generated and potentially reinforce societal stereotypes [84, 85]. Thus, it is necessary to develop more general evaluation metrics and mitigation strategies to prevent the growth of biases in models.

Table 3. Research on Fairness of Vision-language tasks.

| *Ref.* | *Task* | *Method* | *Advantage* | *Limitations* |
|---|---|---|---|---|
| [86] | Image Captioning | Analyzes model behavior by protected characteristics like religion | Improve the understanding of representational issues in captioning | Creating fair VC measurement methods |
| [87] | Image Captioning | A framework using Multi-modal LLMs | Generates culturally-aware captions | Did not evaluate the ethical aspects |
| [77] | Visual Dialogue | Using computational learning theory | Ensure both fairness and human likeness | Not enough human annotations for completely Gender debasing |
| [78] | Visual Question Answering | An evaluation framework for demographic biases in real life | Detailed evaluation of Visual Fairness on LVLMs | Dataset limitations in capturing real world attributes |
| [88] | Visual Question Answering | Developing balanced visual-textual dataset | Reducing language priors and also being explainable | Limited ability to understand visual nuances |
| [89] | Visual Question Answering | Using both modalities | Reducing Unimodal bias in VQA Models | Inherent biases in real world |
| [90] | Visual Question Answering | Counterfactual Samples | Boosting visual-explainable | Models relay in linguistic correlations |

*3)* **Ethical Implications in Visual Dialogue**

The ethical implications of Visual Dialogue systems impact human-computer interaction by presenting challenges related to privacy, data quality, transparency [32]. In intelligence analysis, these systems must handle sensitive data while ensuring fairness and avoiding discrimination. Visual analytics, which merges machine learning with interactive visual interfaces, is essential for addressing these ethical issues by helping analysts interpret complex data and fostering trust, and knowledge generation [93].

Additionally, the development of Multi-Agent Systems for the ethical monitoring of dialogue systems highlights the need for these tools to be built on ethical principles [94]. Thus, while visual dialogue can enhance empathy and understanding, it also necessitates careful consideration of ethical practices in representation and interaction [95]. Table.4. provides a comparison of recent researches about Ethical Implications.

## III. Discussion

Combining visual and linguistic data enhances intuitive interaction by aligning human perception with machine understanding. Vision-language research seeks to effectively integrate Computer Vision and NLP. This study tackles vision-language tasks by addressing the key challenges modern AI systems encounter.

● **Explainability**. In Vision-language tasks, particularly in Visual Dialogue systems, effective communication and collaboration between humans and AI systems significantly depend on the clarity of the AI's explanations. For example, deconfounded learning, outlined in [63, 66, 96], improves vision-language explanations and incorporates interactive mechanisms to boost user feedback and system performance.

Natural Language Explanations (NLEs) are especially advantageous. They provide human-friendly insights into AI Decision-making, making complex processes more accessible [97]. Importantly, these studies indicate that 'helpful' explanations can significantly improve performance on Vision-language tasks. This underscores the importance of incorporating effective explanation mechanisms in AI systems to support user understanding and improve decision-making in Vision-language tasks.

● **Fairness**. The advancement of AI has also encountered critical Fairness issues [77, 79, 90]. Large Vision-language models, such as CLIP [98] variants, inherit many gender biases [99, 100]. Therefore, when we leverage the potential of these models for downstream tasks (such as VQA, Image captioning, and visual dialogue), labeling a model as 'better" based solely on its higher accuracy in a specific evaluation can be misleading and potentially harmful [84]. Imbalanced gender representation in AI datasets exacerbates these problems, leading to biased model predictions.

Our research indicates that racial biases influence the analyzed tasks, as the skin color of depicted individuals notably influences their performance and word choices. These results highlight the necessity for improved data representation and training methods to reduce bias, ensuring that AI systems deliver fair and accurate outcomes across all demographic groups. It is highly advisable to explore innovative methods for estimating biases associated with group identities that are not immediately visible. Additionally, examining the influence of representatives of various social groups on these biases can yield profound insights.

Table 4. Research on Ethical Implications of Vision-language tasks.

| Ref. | Task | Method | Advantages | Limitations |
|---|---|---|---|---|
| [101] | Visual Dialogue | Promoting reader engagement via graphic stories. | Graphic narratives foster empathy and ethics in readers | Limited Generalization |
| [51] | Visual Question Answering | To assist medical professionals with unpredictable questions | VQA enables real-time medical QA using unseen images | Medical ML require adaptable problem-solving skills |
| [84] | Image Captioning | Proposed a hybrid metric to mitigate gender biases | Mitigate gender bias and correlations | Analyzing implications of metric biases in real world |
| [83] | All | Propose measures to mitigate bias in language | Emphasize local community needs | Bias towards English |
| [94] | Visual Dialogue | ASP for knowledge representation. LLP for learning ASP rules | Ethical evaluation approach in pilot system | scarcity of training datasets |
| [93] | All | Analyze VA methods for addressing ethics | Scenario-based stakeholder analysis of actors and roles | Training gaps among users for understanding VA systems |
| [85] | Image Captioning | Dataset collection instructions enhancement | Analysis of biases in the COCO dataset for VC | Social biases in VC due to racial and gender |

● **Ethical Implications.** Efforts to mitigate bias in AI are increasing with strategies such as data preprocessing, model selection, and post-processing. Although these methods are successful, they have limitations and raise ethical concerns. (See Table 4.)

Moreover, ethical implications should guide the development of AI systems and promote transparency in the decision-making processes [102]. However, aligning these components remains challenging because focusing on one aspect may negatively affect another. Although a unified framework is theoretically feasible, its practical implementation necessitates careful attention to interdependencies for a balanced approach.

Table 5 shows the evaluation of the effectiveness of the research for each challenge. It compares the approaches introduced for each challenge across tasks, including those applicable to multiple tasks. Research shows that the relationship between fairness and explainability is complex, with these objectives often being independent and not mutually reinforcing when optimized separately [103]. Integrating explainability and fairness into a unified framework is essential for responsible AI deployment.

Studies indicate that there is no single model that is ideal for every situation. It is important to understand how different aspects of systems interact. It has been shown that while improving certain aspects, such as explainability and fairness, may lead to a decline in others [104], like accuracy, some aspects can also support each other. For example, enhancing a model's explainability results in more transparent models. Future research should connect technological capabilities with trustworthiness considerations to create more effective AI systems in real-world applications.

## IV. Future opportunities

Given the rapid growth of Vision Language research, our review is not exhaustive. We focus on the trustworthiness of Vision-language-tasks to provide a comprehensive overview. To further ensure the responsible and ethical deployment of vision-language research, we outline key opportunities for future study.

● **Real-world applications of Trustworthiness principles**

In commercial artificial intelligence models, implementing fairness, explainability, and ethics often faces practical challenges. One important issue is the management of dynamic and interconnected data structures, which the current literature on fairness does not address. Additionally, while explainability techniques are valuable, they may not always be practical for real-time applications due to computational limitations [102]. Despite these challenges, maintaining a commitment to developing fair, explainable, and ethical AI remains a top priority to build trust with users.

● **Advancements in Multimodal Large Language Models (MLLMs)**

Multimodal large language models (MLLMs) have shown remarkable capabilities in capturing complex linguistic and semantic relationships, effectively linking visual and textual elements. These models are typically trained on paired image-text datasets, allowing them to associate visual content with descriptive language. Techniques such as Multi-instance Visual Prompt Generators [105] and dynamic visual projection mechanisms [5] have further enhanced their ability to integrate visual information into language models. Future research should focus on refining these methods to improve multimodal representations [106] and expand their applicability across diverse vision-language tasks.

● **Mitigating Hallucinations in Large Vision-language Models (LVLMs)**

A critical challenge in large vision-language models (LVLMs) is the issue of hallucinations, where models generate incorrect or unfounded content [107]. While approaches like causal hallucination probing [108] have been proposed to address this issue, more robust solutions are needed to effectively reduce hallucinations across various LVLMs. Addressing this issue is essential for improving the reliability and performance of these models in real-world applications, ensuring they produce accurate and trustworthy outputs.

● **Ensuring Consistency across Tasks**

Consistency across tasks is important for trustworthiness in vision-language systems. Inconsistencies in model behavior can undermine user trust and hinder integration into larger systems. Future research should prioritize the development of better benchmarks and training methodologies to enhance model reliability across diverse tasks and domains. This includes creating standardized evaluation frameworks that consider fairness, transparency, and ethical considerations, ensuring models perform consistently and reliably in various contexts.

## V. Conclusion

This study reviews recent research on transparency, fairness, and ethical considerations in key vision-language tasks, including Visual Question Answering (VQA), Image Captioning, and Visual Dialogue. Such research is essential for developing trustworthy multimodal AI systems. Although significant progress has been made in ensuring fairness in VQA, we recommend that future researchers focus on mitigating bias in large language models (LLMs), given their widespread use.

Additionally, this study underscores the considerable advancements in addressing ethical considerations in image captioning. With the rise of large vision language models (LVLMs) and multimodal large language models (MLLMs), the future of vision-language tasks appears promising. As these systems continue to advance, it is essential to embed ethical guidelines and transparency mechanisms in their design and training. Collaborative efforts among researchers, policymakers, and industry stakeholders will be important to achieving these objectives.

Table 5. An overview of research on three key Vision-language tasks focusing on three important principles of Trustworthiness.

| *Task* | *Fairness* | *Transparency* | *Ethics.* |
|---|---|---|---|
| VQA | High | Medium | Low |
| Visual Dialogue | Medium | Medium | Medium |
| Image Captioning | Medium | Medium | High |

**Declarations**

*Funding*
This research did not receive any grant from funding agencies in the public, commercial, or non-profit sectors.

*Authors' contributions*
MS: Study design, revision of the manuscript, interpretation of the results, drafting the manuscript; AT: Study design, conceptualization, Supervision, revision of the manuscript.

*Conflict of interest*
The authors declare that there is no conflict of interest.

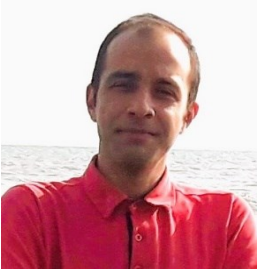

**Mohammad Saleh** holds a Bachelor's degree in Software Engineering from Tehran South University and a Master's degree in Data Science, specializing in Vision-Language Models from the University of Science and Culture. His research interests encompass Computer Vision, Deep Generative Models, and Trustworthiness. He investigates the effects of emerging technologies, including explainable AI, reasoning AI, and diffusion models, on Computer Vision algorithms.

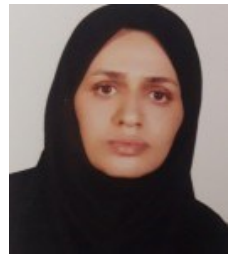

**Azadeh Tabatabaei** earned her M.Sc. in Computer Science from Amirkabir University of Technology in Tehran, Iran, in 2009, and obtained her Ph.D. in Computer Engineering from Sharif University of Technology in 2016. With over a decade of academic experience, her research focuses on deep learning and generative models. She is fascinated by the potential of these technologies to tackle complex challenges and transform industries, including healthcare and finance. She is a faculty member at the University of Science and Culture.